\documentclass[9pt,twocolumn,twoside]{osajnl}

\journal{ao} 

\setboolean{shortarticle}{false} 

\usepackage{physics}

\title{A novel method for measuring the resonant absorption coefficient of rare-earth-doped optical fibers}

\author[1,2]{Mostafa Peysokhan}
\author[1,2]{Esmaeil Mobini}
\author[1,2]{Behnam Abaie}
\author[1,2,*]{Arash Mafi}

\affil[1]{Department of Physics \& Astronomy, University of New Mexico, Albuquerque, NM 87131, USA}
\affil[2]{Center for High Technology Materials, University of New Mexico, Albuquerque, NM 87106, USA}

\affil[*]{Corresponding author: mafi@unm.edu}

\ociscodes{(230.2285) Fiber devices and optical amplifiers; (140.3615) Lasers, ytterbium; (160.5690) Rare-earth-doped materials; (120.4530) Optical constants.}

\begin{abstract}
A non-destructive method for measuring the resonant absorption coefficient of rare-earth-doped optical fibers is introduced.
It can be applied to a broad range of fiber designs and host materials. The method compares the side-collected 
spontaneous emission at two arbitrary locations along the fiber as a function of 
the pump wavelength to extract the absorption coefficient. It provides an attractive and accurate alternative 
to other available techniques. In particular, the proposed method is superior to the cut-back method, which destroys the sample and is 
prone to inaccuracies due to the cladding mode contamination. Moreover, because it does not involve any mechanical movement,
it can be used for fragile optical fibers.
\end{abstract}

\setboolean{displaycopyright}{true}

\begin{document}
\maketitle
\section{Introduction}
Fiber lasers and amplifiers are widely adopted in industry and scientific research because of their high power, good beam quality, 
and ease of operation~\cite{richardson2010high,zervas2014high}. In order to design and optimize fiber lasers and amplifiers, it is essential to
know the geometrical and optical properties of the optical fiber gain medium to a high degree of accuracy~\cite{kouznetsov2002efficiency,Yahel:03,Mafi04,Yahel:06}. 
Such characteristics may be considerably different from those anticipated from the fiber preform and can be altered 
during the fiber drawing process. Therefore, it is important to accurately measure these characteristics directly in the fiber. 
An important property of a rare-earth-doped optical fiber is the resonant absorption coefficient $\alpha_r(\lambda)$, 
which can be determined from the dopant density $N_0$
and the absorption cross section $\sigma_{\rm abs}(\lambda)$. However, $\sigma_{\rm abs}(\lambda)$ is strongly dependent on the host glass,
which can be affected during the preform fabrication and drawing. The dopant density profile can also be modified during the fiber drawing
because of diffusion; therefore, it is imperative to determine $\alpha_r(\lambda)$ directly using the optical fiber.

In this work, we present a novel method that can be used to accurately determine $\alpha_r(\lambda)$ in the
presence of rare-earth ions in an optical fiber at all relevant wavelengths.
The method is based on analyzing the emitted side-light, which contains both fluorescence and pump scattering at different locations
along the fiber. It is a universal technique that can be applied to single-mode, multi-mode, large-mode-area,
photonic crystal, and double-clad rare-earth-doped optical fibers. It is also applicable to fibers made from different materials such as 
ZBLAN, silica, or chalcogenides. Because the method does not involve the movement of any mechanical or optical components during 
the measurement process, it can be readily applied to fragile fibers~\cite{zhu2010high}, including highly tapered fibers~\cite{Kerttula:12}.

The presented method is an alternative to the cut-back method, which is widely used to measure the absorption 
coefficient of optical fibers~\cite{pask1995ytterbium}.
In the cut-back method, the output power from the fiber is measured by gradually cutting back the fiber from the end and reducing its length~\cite{pask1995ytterbium}. 
The cut-back method is destructive; therefore, it cannot be employed in experiments that need to be performed
on a single piece of optical fiber. In a sensitive experiment, e.g. for laser cooling, even a slight sample-to-sample 
variation can affect the outcome; therefore, two pieces of the same fiber may not perform the same way and 
must be characterized individually~\cite{mungan1997laser,gosnell1999laser,melgaard2014identification}.
Another issue involves the excitation of the cladding modes that contaminate the  cut-back measurements in short pieces of the fiber~\cite{yao2009low}.
Moreover, in the cut-back measurements of highly absorbing rare-earth-doped optical fibers, because the core must be pumped well below the
saturation intensity, the output signal can be quite weak and even comparable to the cladding power contamination.
Finally, the cut-back method involves undesirable mechanical processing such as cleaving, polishing, and inspecting the fiber 
that at best can be quite elaborate, and in cases involving fragile fibers totally impractical.

We already mentioned that our proposed method is highly advantageous for characterizing fibers for laser cooling. In a similar context,
the accurate determination of $\alpha_r(\lambda)$ is essential for designing radiation-balanced lasers 
(RBLs)~\cite{bowman1999lasers,bowman1999radiation, nemova2009athermal,Nemova:11,bowman2010minimizing,bowman2016low,peysokhan2017minimizing,yang2018radiation,mobini2018thermal, mobini2018radiation,KnallIEEE}. 
RBLs have been proposed as a way to mitigate the thermal issues in high-power fiber lasers, which have hindered the progress in power-scaling 
because of the thermally induced transverse mode 
instability~\cite{brown2001thermal,Li:05,eidam2011experimental,ward2012origin,jauregui2012physical}. 
RBLs operate based on the fluorescence cooling principle, 
in which the rare-earth-doped optical fiber is pumped at a wavelength, which is higher than the mean fluorescence wavelength of the active ions; 
therefore, the anti-Stokes fluorescence removes some of the excess heat~\cite{bowman1999lasers}. In RBLs, the 
heat generated due to the quantum defect, parasitic background absorption of the pump and laser, and the non-radiative relaxation of the excited rare-earth ions
is balanced against the fluorescence cooling. RBLs pose stringent requirements on the types and levels of dopants, as well as the host materials. In particular, 
the parasitic background absorption ($\alpha_b$) must be quite small for RBLs to work. Our proposed method, when combined with the laser-induced temperature 
modulation spectrum (LITMoS) test developed in Sheik-Bahae's research group~\cite{melgaard2014identification}, allows us to also accurately 
determine $\alpha_b$ for the doped fiber and the cooling efficiency of rare-earth doped fibers~\cite{mungan1997laser, gosnell1999laser}.

\section{Theory}
We refer to this new technique as ``measuring the absorption coefficient via side-light analysis'' (MACSLA).
This method is based on the fact that when a rare-earth doped optical fiber is pumped far below the saturation intensity, 
the spontaneous emission power emitted from the side of the fiber is
directly proportional to the pump power. The method compares the side-collected spontaneous emission at two 
arbitrary locations along the fiber as a function of the pump wavelength and employs the McCumber theory~\cite{mccumber1964einstein}
to extract the spectral form of the absorption coefficient.

\begin{figure}[h]
\includegraphics[width=3.4in]{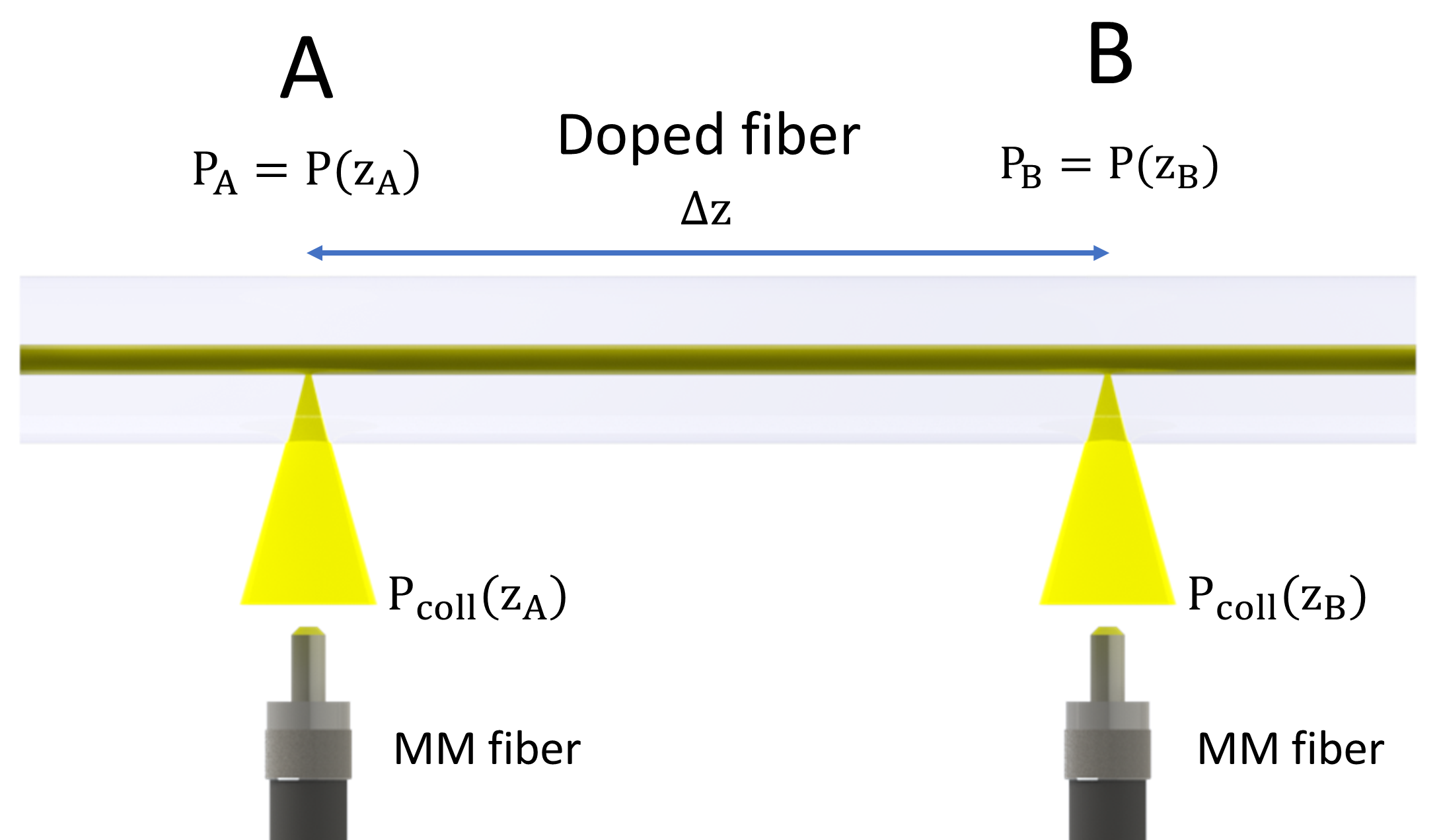}
\caption{Schematic of the propagation of the pump power in the optical fiber and the collection of the spontaneous emission from the side of the rare-earth-doped optical fiber.}
\label{fig:fiber}
\end{figure}
Figure~\ref{fig:fiber} shows a schematic of the proposed method. The pump propagates through the core of the optical fiber from 
left to right. The pump wavelength is assumed to be in the proximity of the peak absorption wavelength such that $\alpha_r(\lambda)$ 
is much larger than $\alpha_b$. The pump intensity in the fiber core is assumed to be far below the saturation intensity; 
therefore, the pump power propagating in the core, $P_{\rm core}(z)$, attenuates exponentially due to the absorption by the rare-earth dopants:
\begin{align}
P_{\rm core}(z) = P_0\,\exp\left(-\alpha_r (\lambda)\,z\right),
\label{Eq:pz}
\end{align}
where $P_0$ is input pump power in the core at $z=0$.

The side-emitted spontaneous emission power is collected by two large-core high-numerical-aperture multimode optical fibers at points {\bf A} and {\bf B} 
along the fiber, which are separated by a distance $\Delta z$. The collection points {\bf A} and {\bf B} and their distance $\Delta z$ remain unchanged
through the experiment.
The collection efficiencies of the two multimode fibers may be slightly different due to 
inevitable misalignments. Therefore, we can write 
\begin{subequations}
\begin{align}
\label{Eq:PAPBA}
&P_{\rm coll}(z_A)\,=\,\gamma_A\,P_{\rm core}(z_A),\\
\label{Eq:PAPBB}
&P_{\rm coll}(z_B)\,=\,\gamma_B\,P_{\rm core}(z_B),
\end{align}
\end{subequations}
where $P_{\rm coll}(z_A)$ and $P_{\rm coll}(z_B)$ are the collected powers at points {\bf A} and {\bf B}, respectively.
$\gamma_A$ and $\gamma_B$ are coefficients that relate the propagating power in the core to the collected spontaneous 
emission power, which also incorporate the coupling efficiencies to the multimode fibers at points {\bf A} and {\bf B}, respectively.
We now divide Eq.~\ref{Eq:PAPBB} by Eq.~\ref{Eq:PAPBA}, take the natural logarithm of both sides, and obtain:
\begin{align}
\label{Eq:ratio}
r(\lambda)\,=\,
\ln\left(\gamma_B/\gamma_A\right)-\alpha_r(\lambda)\,\Delta z,
\end{align}
where 
\begin{align}
\label{Eq:ratio2}
r(\lambda)\,=\,\ln\left(\dfrac{P_{\rm coll}(z_B)}{P_{\rm coll}(z_A)}\right).
\end{align}

In Eq.~\ref{Eq:ratio}, $\alpha_r(\lambda)$ follows a strict spectral function of the form (see Appendix A):
\begin{align}
\alpha_r(\lambda) \propto \lambda^5\, S(\lambda)\, \exp\left(\dfrac{hc}{\lambda k_B T}\right), 
\label{Eq:abs}
\end{align}
where $S(\lambda)$ is the emission power spectral density measured by the optical spectrum analyzer, $h$ is the Planck constant, $k_B$ is
the Boltzmann constant, and $c$ is the speed of light in vacuum.
We also assume that the ratio $\gamma_B/\gamma_A$ is wavelength independent over the narrow range of wavelengths used in this experiment.
Therefore, the left-side in Eq.~\ref{Eq:ratio}, $r(\lambda)$, must also follow the spectral form in Eq.~\ref{Eq:abs} when the pump 
wavelength is varied. Because the spectral shape of 
$\alpha_r(\lambda)$ is obtained from Eq.~\ref{Eq:abs}, all that is needed is to find its overall magnitude by balancing 
the left-side and right-side in Eq.~\ref{Eq:ratio} over the respective wavelengths. Therefore, we replace $\alpha_r(\lambda)$ 
in Eq.~\ref{Eq:ratio} with $\alpha^p_r\times \widetilde{\alpha}_r(\lambda)$, where $\widetilde{\alpha}_r(\lambda)$
is the absorption coefficient normalized to its peak value, $\alpha^p_r=\alpha_r(\lambda_{\rm peak})$. 
This way, we can determine both $\gamma_B/\gamma_A$ and $\alpha^p_r$ through a fitting procedure that involves measurements of $r(\lambda)$
and $\widetilde{\alpha}_r(\lambda)$ at multiple wavelengths near the peak absorption wavelength.

\section{Experiment}
In our experiment, we used a commercial Yb-doped optical fiber (SM-YSF-LO-HP, Nufern, Inc.) to demonstrate the utility of the MACSLA method.
SM-YSF-LO-HP is a low-doped Yb-silica single-mode and single-clad optical fiber. As we mentioned in the previous section,
in order to use the Beer-Lambert exponential decay form in Eq.~\ref{Eq:pz}, the pump intensity must be kept 
considerably below the saturation intensity. As such, we first measured the pump saturation power ($P_{\rm sat}$) by pumping the 
core of the doped fiber ($P_{\rm core}$) and measuring the side spontaneous emission power ($P_{\rm spont}$) for different values of the pump power 
at 976\,nm wavelength. The measurements were fitted to the functional form of the saturated power in a doped fiber~\cite{zech1995measurement}
\begin{align}
P_{\rm spont}(P_{\rm core}) \propto \frac{P_{\rm core}}{1+P_{\rm core}/P_{\rm sat}}.
\label{Eq:spont}
\end{align}
For our fiber, the saturation power was determined to be 966\,\textmu W. In our later experiments, $P_{\rm core}$ was kept 
below 5\% of the saturation power to make sure that Eq.~\ref{Eq:pz} could be reasonably applied (see Appendix B).

\begin{figure}[h]
\includegraphics[width=3.3in]{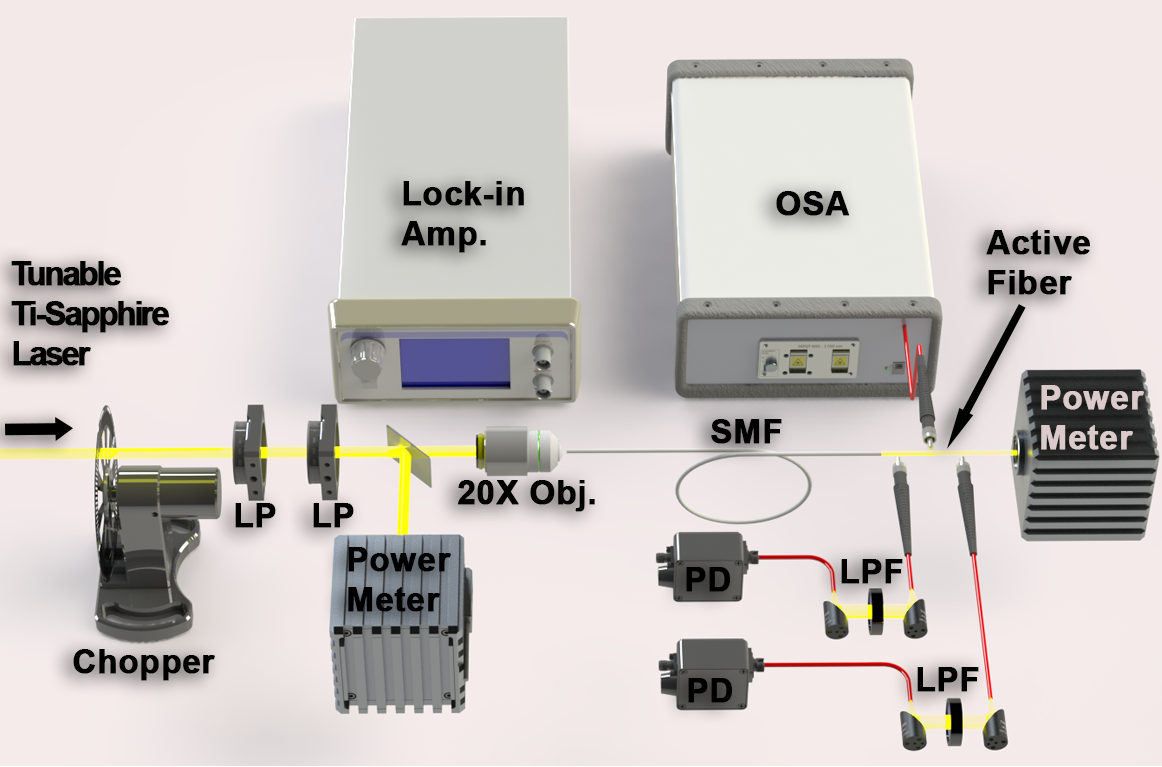}%
\caption{\label{fig:setup} Experimental setup for measuring the absorption coefficient using the MACSLA method. LP stands for linear polarizer,
OSA for optical spectrum analyzer, LPF for long-pass filter, PD for photodetector, and SMF for single-mode fiber.}
\end{figure}
\begin{figure}[ht]
\centering
\includegraphics[width=2.7in]{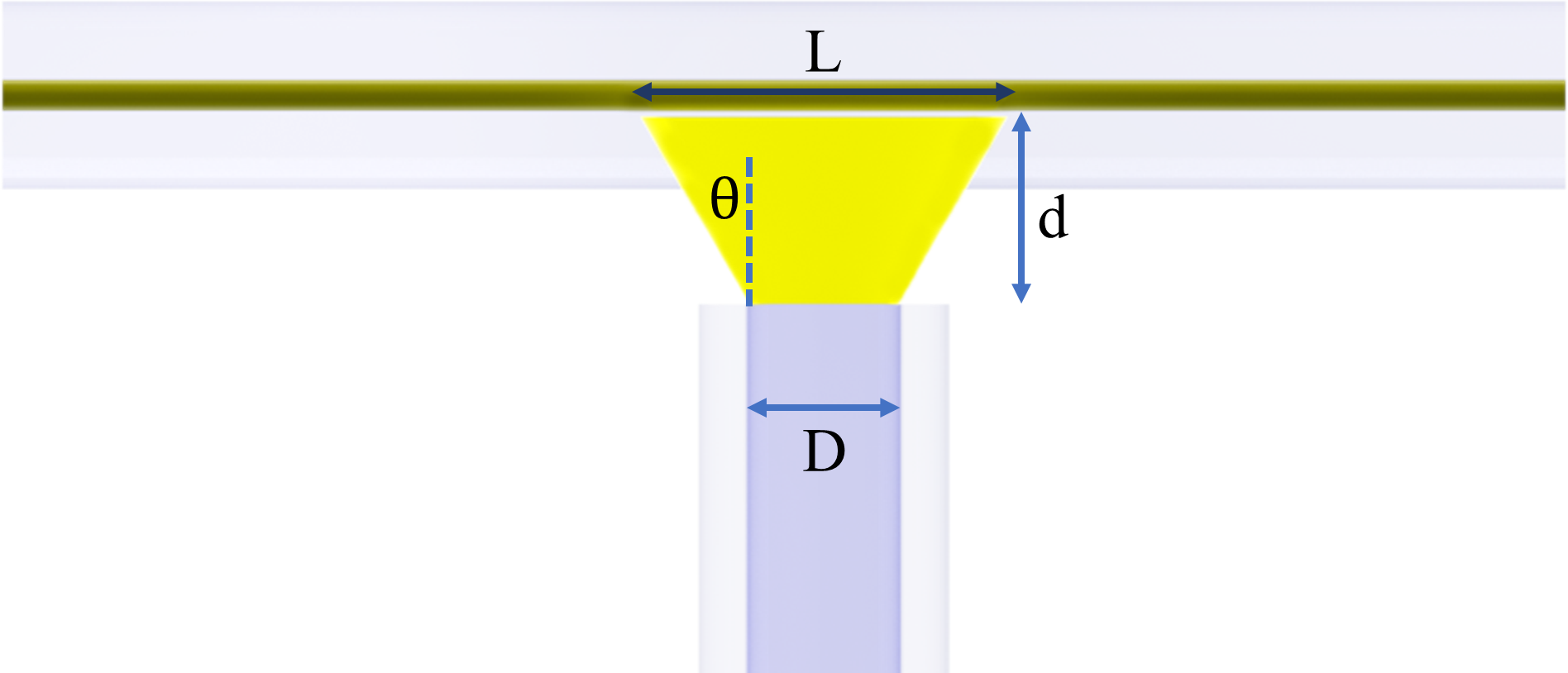}
\caption{Schematic of the fluorescence emission collected by the multimode fiber from the side of the Yb-doped fiber for the estimation of the fiber segment length, $L$, 
from which the side spontaneous emission is collected (see Eq.~\ref{Eq:point}).}
\label{fig:side}
\end{figure}

The experimental setup that is used in MACSLA for measuring the absorption coefficient is shown in Fig.~\ref{fig:setup}. 
The fiber was pumped by a tunable continuous-wave (CW) Ti:Sapphire laser and the side spontaneous emission power was collected at points {\bf A} and {\bf B}
using two multimode fibers (M124L02, Thorlabs, Inc.), which were connected to InGaAs Detectors (DET08CFC, Thorlabs, Inc.).
Long-pass filters ($\ge$1\textmu m) were placed between the multimode fibers and the detectors in order to 
ensure that scattered pump does not contaminate the fluorescence signal.
The distance between points {\bf A} and {\bf B}, $\Delta z$, must be substantially larger than the typical length of the 
Yb-doped fiber segment, $L$, from which the side spontaneous emission is collected by each light-collecting multimode fiber. This is to ensure 
that the collection segment can be considered a point for all practical purposes relative to $\Delta z$. 
The length of the fiber segment is approximately given by
\begin{align}
L\,=\,D+2d\tan\theta,\qquad \sin\theta={\rm NA},
\label{Eq:point}
\end{align}
where $D$ is the diameter of the core of the multimode fiber, $d$ is the distance between the input 
tip of the light-collecting multimode fiber and the core-cladding interface of the Yb-doped fiber, $\theta$ is the maximum acceptance angle of the 
light-collecting  multimode fiber, and ${\rm NA}$ is the numerical aperture of the light-collecting multimode fiber, as shown in Fig.~\ref{fig:side}. 
Using typical values for our experiment in Eq.~\ref{Eq:point}, one concludes that $L$ is on the order of 0.5mm and $\Delta z\gg 0.5{\rm mm}$ is required.
Here, we have used $D=400$\textmu m, ${\rm NA}=0.5$, and $d\approx 100$\textmu m, including approximately 40\textmu m of gap between the cladding of the doped fiber
and the tip of the light-collecting  multimode fiber.
In our experiments, the light-collecting multimode fibers nearly touched the cladding of the Yb-doped fiber, so $d$ was comparable to the radius of the 
Yb-doped fiber. 
At the same time, $\Delta z$ must be sufficiently large to allow for a distinct nontrivial collected power ratio in the form of $r(\lambda)$, 
while it must be small enough to ensure that the signal to noise ratio at point {\bf B} is not much different from that of point {\bf A},
because the signal drops exponentially with $\Delta z$.
For our experiment, we chose $\Delta z\,=\,3.23\,{\rm cm}$, which satisfied all these constraints.

The detectors were connected to a lock-in amplifier that was used to extract the weak spontaneous emission signal from the
noisy background. In order to use the lock-in amplifier, the pump was modulated at 1\,KHz with a commercial chopper.
Two linear polarizers were used to attenuate the pump power and keep the core power far below the saturation.
A beam-splitter and a power-meter were used to measure the input power, and another power-meter was placed at the end of the doped 
fiber to monitor the output power as a secondary check to make sure that the fiber core power remained far below saturation throughout the experiment.
The pump power was coupled to a passive single-mode fiber by a 20X microscope objective, and the single-mode fiber 
was fusion-spliced to the doped fiber to deliver the pump power. As the pump wavelength is varied, the input pump power changes
slightly; however, the MACSLA method relies only on the power ratios collected at points {\bf A} and {\bf B} and is not affected by such power
variations.

\begin{figure}[h]
\includegraphics[width=3.4in]{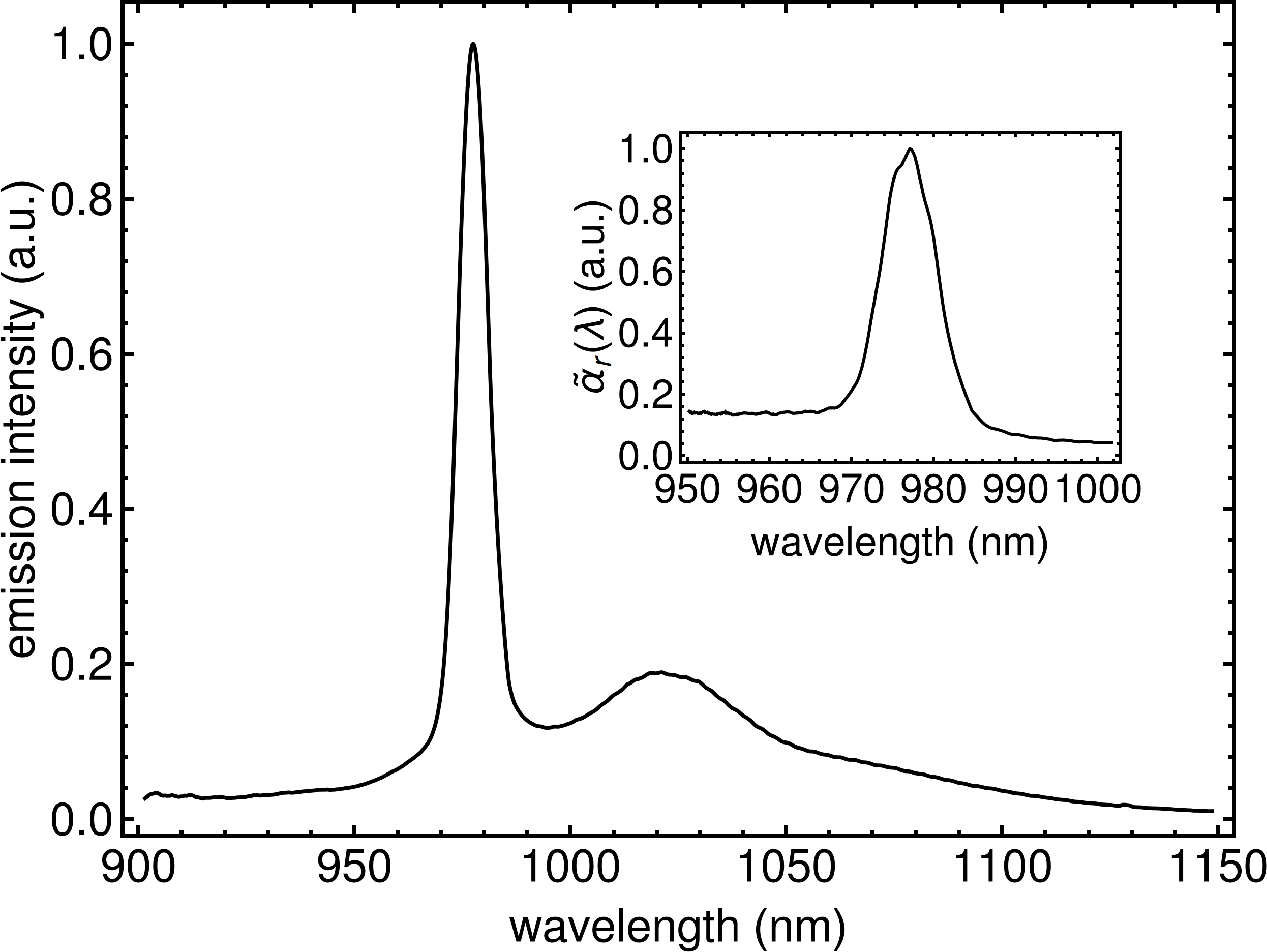}
\caption{\label{fig:spectrum} The emission power spectral density $S(\lambda)$, which is measured by the optical spectrum analyzer is plotted
in arbitrary units. The inset shows the resonant absorption coefficient, which is normalized to its peak value, and is calculated by using the 
McCumber theory~\cite{mccumber1964einstein}.}
\end{figure}

For the fitting procedure, we chose seven different pump wavelengths near the absorption peak wavelength for the Yb-silica 
fiber ($\lambda_{\rm peak}$\,=\,977\,nm) by tuning the operating 
wavelength of the CW Ti:Sapphire laser. For each wavelength, the emission signal power was measured at positions {\bf A} and {\bf B}
over sufficient time windows until the desired signal-to-noise-ratio was achieved and the error-bars were obtained from the lock-in amplifier.
The distance between points {\bf A} and {\bf B} was also measured by a digital caliper.
The power spectral density $S(\lambda)$ of the Yb-silica fiber is shown in Fig.~\ref{fig:spectrum}. The inset shows the
resonant absorption coefficient, which is normalized to its peak value, and is calculated by using the McCumber theory~\cite{mccumber1964einstein}.
\begin{figure}[htbp]
\includegraphics[width=3.3in]{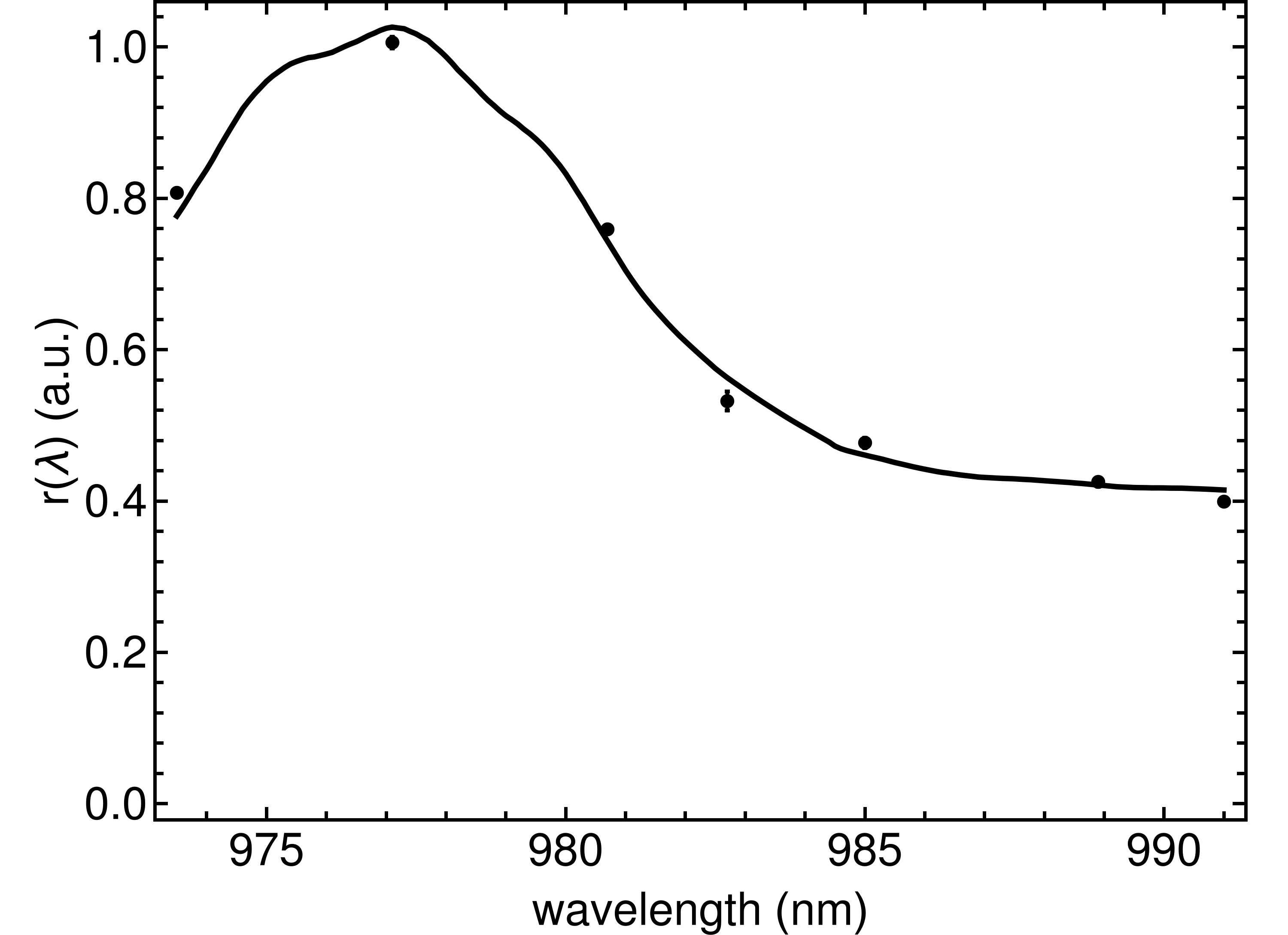}
\caption{\label{fig:final} The points with error-bars indicate the values of $r(\lambda)$ from Eq.~\ref{Eq:ratio2} measured at seven different wavelengths
near the peak of the resonant absorption coefficient. The fitting curve comes directly from the
resonant absorption spectrum shown as the inset in Fig.~\ref{fig:spectrum}. The fitting parameters are $\alpha^p_r$ and $\gamma_B/\gamma_A$.}
\end{figure}

\section{Results and Discussion}
The fitted line over the experimental measurements related to Eq.~\ref{Eq:ratio} are shown in Fig.~\ref{fig:final}. The points (with error-bars) 
indicate the values of $r(\lambda)$ measured at seven different wavelengths, and the fitting curve comes directly from the
resonant absorption spectrum shown as the inset in Fig.~\ref{fig:spectrum}. The outcome of the fitting procedure was the peak value of the resonant
absorption coefficient $\alpha^p_r\,=\,0.198\pm 0.008\,{\rm cm}^{-1}$. While unimportant to the procedure, the fitting also resulted in $\gamma_B/\gamma_A\,=\,0.69$.  
Our result for $\alpha^p_r$ should be compared with the value reported by the vendor, which is $0.220 \pm 0.033\,{\rm cm}^{-1}$. 
We also performed the cut-back as a point of comparison for measuring the peak value of the resonant absorption coefficient.
We used the CW Ti:Sapphire laser as the pump operating at the peak resonant absorption wavelength of the Yb-doped fiber
and coupled its output using an objective lens to a 50\,${\rm cm}$ segment of a passive single-mode
fiber (HP-980, Nufern), while the other end the single-mode fiber was fusion-spliced to the Yb-doped fiber.

The intermediate step of using a passive single-mode fiber to couple light to the Yb-doped fiber provided two main advantages: first, 
it reduced the amount of pump coupling to the cladding modes of the gain fiber for enhanced accuracy; and second, it allowed for the inevitable 
moving and bending as a result of the incremental cleaving of the gain fiber without having to worry about the fiber alignment to the pump.
We also applied index matching gel on the cladding of the Yb-doped fiber to reduced cladding light contamination of our measurement. Our careful
cut-back method measurement of the peak value of the resonant  absorption coefficient resulted in $\alpha^p_r\,=\,0.203\,{\rm cm}^{-1}$, 
which is closer to the value provided
by the MACSLA method (see also Appendix B for potential corrections to the result from MACSLA).
While the experimental results reported here are for a single segment of the Yb-doped silica fiber, we performed the same procedure on 
a segment of a Yb-doped ZBLAN fiber and the results came in agreement with those reported by the vendor. We also note that the results are
insensitive to the actual value of $\Delta z$--while we reported $\Delta z\,=\,3.23\,{\rm cm}$ in our experiment simply because that was the value
we measured after placing the collection fibers, the results would have been as good if we chose other values such as $\Delta z\,=\,4\,{\rm cm}$.

We would like to comment on a recent pioneering method proposed by Min Oh, et al.~\cite{MinOH}, where they also employ the side-light analysis 
to measure $\alpha_r(\lambda)$. In their procedure, the doped fiber is pumped at a fixed wavelength and the spontaneous emission is measured at
different positions along the fiber by using an optical spectrometer. They measure $\alpha_r$ at the respective wavelength by fitting the 
side-collected power to the Beer-Lambert exponential decay form in Eq.~\ref{Eq:pz}. In their method, the coupling efficiency to the side-collecting 
fiber is assumed to remain unchanged at different locations along the doped fiber. This assumption is likely to result in measurement 
inaccuracies if the gain fiber is single-mode with a small core. The fitting to the Beer-Lambert exponential decay form necessitates mechanical
movement to obtain measurements at multiple points; therefore, optical alignment may become an issue and it will be hard to maintain a
uniform coupling efficiency to the side-collecting fiber at all points.
Moreover, the requirement to
keep the pump power far below the saturation power, which is on the order of 1~mW in single-mode fibers, necessitates high-sensitivity spectrometers for adequate 
signal-to-noise-ratio.

\section{Summary and Conclusion}
In summary, the MACSLA method provides an attractive and accurate alternative to other techniques for measuring the 
resonant absorption coefficient in rare-earth-doped optical fibers. In particular, it is superior to the cut-back method, 
which destroys the sample and is prone to inaccuracies due to the cladding mode contamination. When combined with the 
LITMoS test~\cite{melgaard2014identification,peysokhan2018measuring}, the MACSLA method allows one to also 
determine the parasitic background absorption coefficient ($\alpha_b$). In laser cooling experiments and RBLs, the cooling efficiency is improved 
by reducing the ratio of background absorption coefficient to the resonant absorption 
coefficient ($\alpha_b/\alpha_r$)~\cite{melgaard2014identification,mungan1997laser, gosnell1999laser,mobini2017laser, mobini2018investigation, knall2018model,peysokhan2018measuring}. 
Our techniques enables an accurate determination of $\alpha_b$ that is essential to design and interpret such fluorescence cooling experiments.

One of the main advantages of the MACSLA method is the fact that it does not require 
precise alignments, which makes it suitable for commercial applications. Moreover, the technique does not require an 
accurate knowledge of the actual coupled power into the medium, hence one is not worried about surface reflections and 
scatterings. In practice, the lock-in amplifier may not be required if one uses a high-sensitivity detector such as a 
low threshold avalanche photodiode. We verified this in a separate measurement for a multimode Yb-doped optical fiber,
where the side-collected spontaneous emission signal was stronger than that of a single-mode Yb-doped fiber due to a 
higher value of pump power. Finally, we would like to emphasize that while the MACSLA method is used to extract the 
peak value of the resonant absorption coefficient, when combined with the emission power spectral density $S(\lambda)$, 
which is measured by the optical spectrum analyzer as in Fig.~\ref{fig:spectrum}, provides a full characterization of
the resonant absorption coefficient at all relevant wavelengths and is not limited to the vicinity of the pump wavelength. 
\section*{Appendix A}
In this Appendix, we would like to justify the form of Eq.~\ref{Eq:abs}.
We start with the McCumber theory, which relates the absorption cross section ($\sigma_{\rm abs}(\nu)$) and 
emission cross section ($\sigma_{\rm em}(\nu)$) of dopants in solid-state media~\cite{mccumber1964einstein}:
\begin{align}
\sigma_{\rm abs}(\nu)\,=\,\sigma_{\rm em} (\nu)\exp\left(\dfrac{h\nu - \epsilon}{k_BT}\right).
\label{Eq:sabs} 
\end{align}
$\nu$ is the frequency of light, and $\epsilon$ is the so-called ``zero-line'' energy.
The emission cross section can be formally obtained from~\cite{SalehBook}:
\begin{align}
\sigma_{\rm em} (\nu) = A_{21}\, g(\nu)\, \frac{\lambda^2}{8 \pi n^2},
\label{Eq:sem}
\end{align}
where the $A_{21}$ is the Einstein {\em A-coefficient}, $g(\nu)$  is the normalized lineshape function, 
$\lambda$ is the free space wavelength, and $n$ is the refractive index of the medium. The resonant absorption 
coefficient, $\alpha_r(\lambda)$, can be expressed as
\begin{align}
\alpha_r\,=\, N_1\sigma_{\rm abs}-N_2\sigma_{\rm em}\approx N_0\sigma_{\rm abs},
\label{Eq:alpharN2}
\end{align}
where $N_1$ ($N_2$) is the population density of the lower (upper) manifold of the dopant ions, and $N_0=N_1+N_2$ is the dopant ion
number density. The approximation in the right hand side of Eq.~\ref{Eq:alpharN2} holds when the gain material is pumped well below the
saturation intensity such that $N_2\approx 0$ and $N_1\approx N_0$.
The observed spectral florescence intensity, $I(\nu)$, is given by~\cite{SalehBook}:
\begin {align}
I(\nu)\,=\,A_{21}\, g(\nu)\, h\nu\, N_2\,=\,\dfrac{\lambda^2}{c}S(\lambda),
\label{Eq:intensity}
\end{align}
where $S(\lambda)$ is the emission power spectral density, previously introduced in the main part of the manuscript, and 
the rightmost part of Eq.~\ref{Eq:intensity} comes from noting that $I(\nu)d\nu\equiv S(\lambda)d\lambda$. Combining 
Eqs.~\ref{Eq:sabs}--\ref{Eq:intensity}, we obtain
\begin{align}
\alpha_r(\lambda)\approx \exp\left(\dfrac{- \epsilon}{k_BT}\right)\left(\dfrac{N_0/N_2}{8\pi h c^2n^2}\right) \lambda^5\, S(\lambda)\, 
\exp\left(\dfrac{hc}{\lambda k_B T}\right), 
\label{Eq:newabs}
\end{align}
where the wavelength-dependent part gives Eq.~\ref{Eq:abs}.
\section*{Appendix B}
In this Appendix, we estimate the impact of the finite ratio of the $P_{\rm core}$ to $P_{\rm sat}$ on our results. In the low emission 
signal limit, which is valid in our setup, the pump propagation is described by
\begin{align}
\dfrac{d P_{\rm core}(z)}{dz}\,=\,-\alpha_r(\lambda)\frac{P_{\rm core}}{1+P_{\rm core}/P_{\rm sat}},
\label{Eq:pump-abs}
\end{align}
where $\alpha_r(\lambda)\,=\,N_0\sigma_{\rm abs}(\lambda)$.
The formal solution to Eq.~\ref{Eq:pump-abs} is given by 
\begin{align}
P_{\rm core}(z)=P_{\rm sat}\times {\mathcal W}
\left(
\dfrac{P_0}{P_{\rm sat}}
e^{P_0/P_{\rm sat}}\, e^{-\alpha_r(\lambda) z}
\right),
\label{Eq:pump-abs-sol}
\end{align}
where ${\mathcal W}(y)$ is the Lambert W-function (product logarithm function) defined as the principal solution for ${\mathcal W}$ in $y={\mathcal W}e^{\mathcal W}$. The zeroth order term in Taylor expansion of Eq.~\ref{Eq:pump-abs-sol} in $P_0/P_{\rm sat}$ gives Eq.~\ref{Eq:pz}.

If we calculate $r(\lambda)$ from Eq.~\ref{Eq:ratio2} and keep terms to the first order of Taylor expansion in  $P_0/P_{\rm sat}$, we obtain
\begin{align}
\label{Eq:ratio3}
r(\lambda)\,&=\,
\ln\left(\gamma_B/\gamma_A\right)-\alpha_r(\lambda)\,\Delta z\\
\nonumber
&+\left(P_0/P_{\rm sat}\right)
e^{-\alpha_r(\lambda)z_A}(1-e^{-\alpha_r(\lambda)\Delta z}),
\end{align}
The actual expansion parameter for the correction term in 
Eq.~\ref{Eq:ratio3} relative to Eq.~\ref{Eq:ratio} is the ratio of $P_0\exp(-\alpha_rz_A)/P_{\rm sat}$,
due to the fact that $P_0$ is the input power to the fiber but
$P_0\exp(-\alpha_rz_A)$ is the power at the first point of spectral measurement. Employing the fitting procedure 
using Eq.~\ref{Eq:ratio3} with $P_0\exp(-\alpha_rz_A)/P_{\rm sat}=5\%$
 gives the peak value of the
absorption coefficient at $\alpha^p_r\,=\,0.206\pm 0.008\,{\rm cm}^{-1}$, which must be 
compared with $\alpha^p_r\,=\,0.198\pm 0.008\,{\rm cm}^{-1}$ obtained from fitting with Eq.~\ref{Eq:ratio}, which is a 4\% correction.
Such an upward correction is understandable because the increase in the denominator in Eq.~\ref{Eq:pump-abs} due to a finite 
value of $P_{\rm core}/P_{\rm sat}$ must be compensated by an increase in the numerator in the form of rescaling $\alpha_r(\lambda)$ to a larger value.
\section*{Funding Information}
This material is based upon work supported by the Air Force Office of Scientific Research under award number FA9550-16-1-0362 
titled Multidisciplinary Approaches to Radiation Balanced Lasers (MARBLE).
\section*{Acknowledgment}
The authors would like to thank M. Sheik-Bahae, R. I. Epstein, and A. R. Albrecht for illuminating discussions.
\bigskip
\providecommand{\noopsort}[1]{}\providecommand{\singleletter}[1]{#1}%

\end{document}